\documentclass[aps,prl,reprint,superscriptaddress,showpacs,amsmath,amssymb]{revtex4-2}

\usepackage{graphicx}      
\usepackage{dcolumn}     
\usepackage{bm}          
\usepackage{hyperref}
\hypersetup{colorlinks=true, citecolor=blue, urlcolor=blue, linkcolor=blue}
\makeatletter

\newcommand{\Rmnum}[1]{\expandafter\@slowromancap\romannumeral #1@}
\makeatother

\begin{document}

\title{Regulating pedestal pressure profile by microtearing modes in the EAST tokamak}

\author{J. Shi}
\affiliation{Institute of Plasma Physics, Hefei Institutes of Physical Science, Chinese Academy of Sciences, Hefei 230031, China}
\affiliation{University of Science and Technology of China, Hefei 230026, China}

\author{Y. Sun}
\email{ywsun@ipp.ac.cn}
\affiliation{Institute of Plasma Physics, Hefei Institutes of Physical Science, Chinese Academy of Sciences, Hefei 230031, China}

\author{K. Fan}
\affiliation{School of Physics and Astronomy, Institute of Natural Sciences, and MOE-LSC, Shanghai JiaoTong University, Shanghai 200240, China}

\author{T. Shi}
\affiliation{Institute of Plasma Physics, Hefei Institutes of Physical Science, Chinese Academy of Sciences, Hefei 230031, China}
\author{S. Wang}
\affiliation{University of Science and Technology of China, Hefei 230026, China}
\affiliation{Southwestern Institute of Physics, PO Box 432, Chengdu 610041, China}

\author{H. Zhao}
\affiliation{Institute of Plasma Physics, Hefei Institutes of Physical Science, Chinese Academy of Sciences, Hefei 230031, China}

\author{Q. Zang}
\affiliation{Institute of Plasma Physics, Hefei Institutes of Physical Science, Chinese Academy of Sciences, Hefei 230031, China}
\author{Z. Zhou}
\affiliation{Institute of Plasma Physics, Hefei Institutes of Physical Science, Chinese Academy of Sciences, Hefei 230031, China}
\affiliation{University of Science and Technology of China, Hefei 230026, China}

\author{H. Yang}
\affiliation{Institute of Plasma Physics, Hefei Institutes of Physical Science, Chinese Academy of Sciences, Hefei 230031, China}

\author{X. Wu}
\affiliation{Institute of Plasma Physics, Hefei Institutes of Physical Science, Chinese Academy of Sciences, Hefei 230031, China}

\author{J. Liu}
\affiliation{Institute of Plasma Physics, Hefei Institutes of Physical Science, Chinese Academy of Sciences, Hefei 230031, China}
\affiliation{University of Science and Technology of China, Hefei 230026, China}

\author{H. Nan}
\affiliation{Institute of Plasma Physics, Hefei Institutes of Physical Science, Chinese Academy of Sciences, Hefei 230031, China}
\affiliation{University of Science and Technology of China, Hefei 230026, China}

\author{Q. Ma}
\affiliation{Institute of Plasma Physics, Hefei Institutes of Physical Science, Chinese Academy of Sciences, Hefei 230031, China}

\author{H. Sheng}
\affiliation{Institute of Plasma Physics, Hefei Institutes of Physical Science, Chinese Academy of Sciences, Hefei 230031, China}

\author{J. An}
\affiliation{Institute of Plasma Physics, Hefei Institutes of Physical Science, Chinese Academy of Sciences, Hefei 230031, China}
\affiliation{University of Science and Technology of China, Hefei 230026, China}

\author{J. Xu}
\affiliation{Institute of Plasma Physics, Hefei Institutes of Physical Science, Chinese Academy of Sciences, Hefei 230031, China}
\affiliation{Key laboratory of materials Modification by Beams of the Ministry of Education, School of Physics, Dalian University of Technology, Dalian 116024, China}

\author{S. Gu}
\affiliation{Institute of Plasma Physics, Hefei Institutes of Physical Science, Chinese Academy of Sciences, Hefei 230031, China}

\author{H. Wang}
\affiliation{Institute of Plasma Physics, Hefei Institutes of Physical Science, Chinese Academy of Sciences, Hefei 230031, China}

\author{K. He}
\affiliation{Institute of Plasma Physics, Hefei Institutes of Physical Science, Chinese Academy of Sciences, Hefei 230031, China}

\date{\today}

\begin{abstract}
The pedestal pressure profile plays a critical role in determining the overall confinement of tokamak fusion plasmas. 
This letter reports the first evidence on regulating pedestal pressure profile by microtearing modes (MTMs) via limiting the pedestal gradient, which was thought to be limited by kinetic ballooning modes over the last decades. 
This is observed in an ITER-baseline-like target plasma in the EAST tokamak. 
MTMs arrest the recovery of pedestal electron temperature (and hence pressure) by clamping its gradient at a critical value while leaving the density profile essentially unaffected. 
Global fluid and local gyrokinetic simulations confirm the excitation of MTMs when the pedestal electron temperature gradient exceeds a threshold value in this experiment. 
This finding reveals MTMs as an alternative electromagnetic micro-instability that imposes an additional constraint in determining the pedestal structure in tokamaks.
\end{abstract}
\maketitle

Forming pedestal-like pressure profile with steep gradient near the edge in high-confinement mode (H-mode) scenarios  \cite{Wagner1984} is crucial for enhancing the global plasma confinement and fusion gain in a tokamak reactor \cite{Groebner_2013}. 
This motivated intensive studies on the mechanism in determining pedestal structure in the past decades. 
In contrast to the studies on core turbulence transport induced by drift-wave micro-instabilities, the transport in the pedestal region linked to electromagnetic micro-instabilities was much less understood. 
Kinetic ballooning modes (KBMs) were thought for a long time to be the key micro-instability that clamps the pedestal pressure gradient rigidly setting pedestal width, until the excitation of peeling ballooning modes 
causing collapse that limits the pedestal height, which explains the periodic edge localized mode (ELM) events in H-mode plasmas \cite{Snyder_2009,Snyder2011}.
However, magnetic perturbations observed during pedestal recovery stage before the next ELM crash in the type-I ELMy H-mode in some tokamaks \cite{Perez_2004,Manz_2014,Nelson_2021,ChenJ_2021pop}  have been identified as microtearing modes (MTMs) \cite{Hazeltine_1975,Drake1977,Hassam1980_tear}.
MTMs may dominate electron heat transport in the pedestal \cite{Doerk2011,Hatch_2021,Chen_2023NF},
while KBMs may be in the second stability regime as plasma beta (normalized plasma pressure) increases in large tokamaks such as JET \cite{Saarelma_2013,Hatch_2016}.
Therefore, MTMs may be an alternative micro-instability that sets a critical pedestal gradient, which remains to be verified experimentally. 
In this Letter, we report the direct evidence that MTMs, identified by combined experiment and simulation, regulate the pedestal electron pressure through a critical-gradient mechanism in the EAST tokamak.

\begin{figure}[t]
	\centering
	\includegraphics[width=\columnwidth]{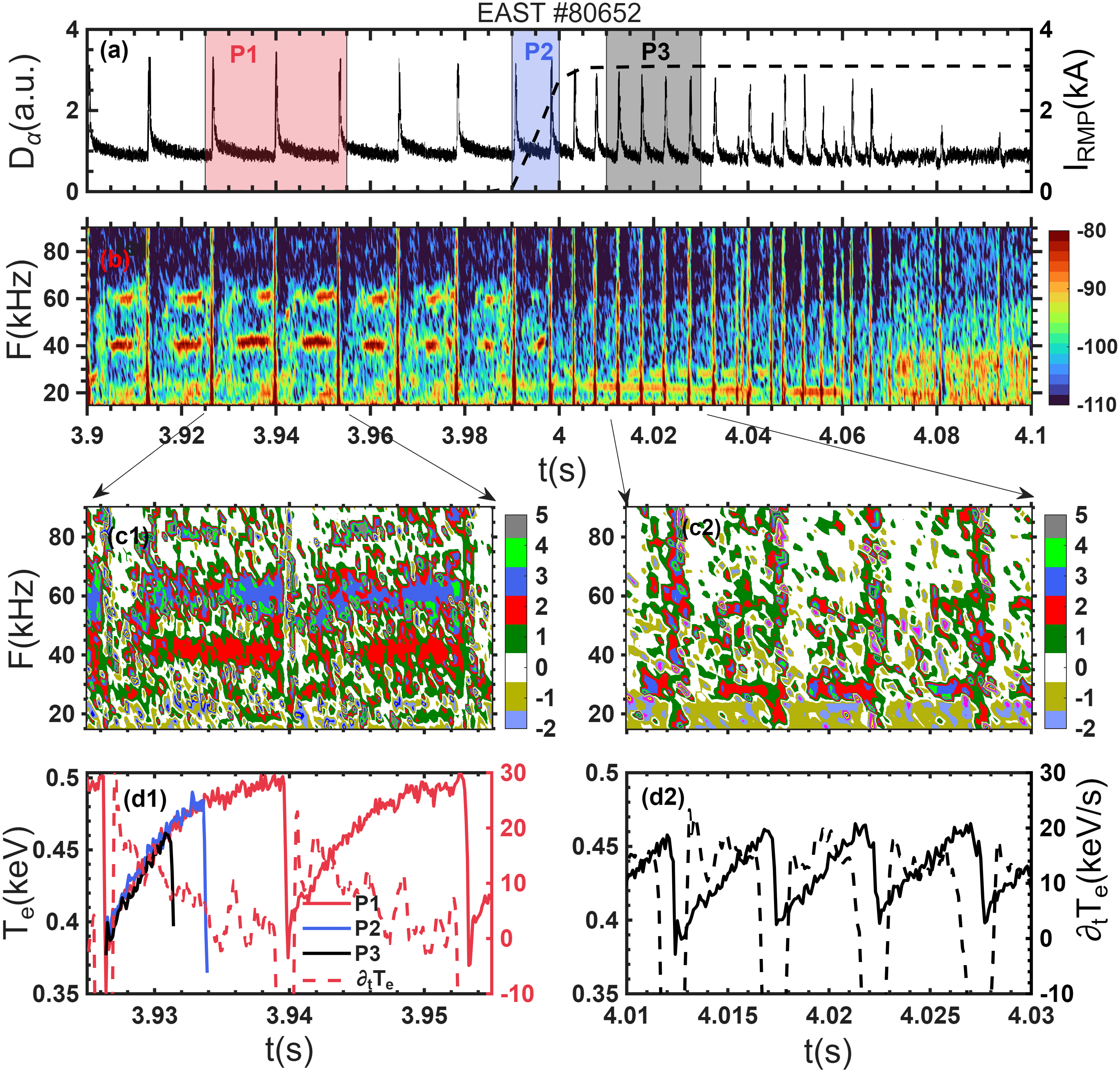}
	\caption{\label{fig:1} 
		Temporal evolution of 
		 (a) divertor D$_\alpha$ emission (solid) and RMP coil current $I_{\mathrm{RMP}}$ (dashed), with three shaded bands marking phases P1-P3; 
		 (b) cross-power spectrum of the magnetic fluctuations,
		together with (c1, c2) toroidal mode number,  (d1, d2)  pedestal-top  $T_{\mathrm{e,ped}}$  (solid) and its time derivative $\mathrm{\partial}_t T_{\mathrm{e,ped}}$ (dashed) in the two zoomed windows P1 and P3  in the EAST discharge \#80652.
		For comparison, the recovery of $T_{\mathrm{e,ped}}$  at P2 (blue) and P3 (black) is also overlaid in (d1) by aligning the onset time.
	}
\end{figure}

This experiment was carried out in type-\Rmnum{1} ELMy H-mode for the ITER-baseline scenario in EAST. 
In this experiment, the toroidal magnetic field is $B_T \simeq 1.6\,\mathrm{T}$, the safety factor at the surface with 95$\%$ of the normalized poloidal magnetic flux is $q_{95} \sim 3.7$,
the normalized beta is $\beta_N\sim1.6$,
and the normalized electron collisionality near the pedestal top is $\nu_{\ast,ped}\sim0.7$.
Fig.~\ref{fig:1} shows one example in the discharge \#80652.  
An $n=4$ Resonant Magnetic Perturbation (RMP) was applied at around $t=$4s. 
Here, $n$ is the toroidal mode number.
It is shown that the recovery of pedestal electron temperature is clearly arrested by discrete low-$n$ modes during the inter-ELM phase in the P1 phase (no RMP).
Magnetic fluctuations measured by Mirnov coils show the appearance of discrete modes with toroidal mode numbers $n=1$--$4$ at approximately $20, 40, 60$ and $80$ kHz, propagating in the electron diamagnetic direction (EDD) (positive $n$ value here).
Here, the $n=2$ and $n=3$ harmonics are the dominant ones. 
The excitation of these modes coincides with the pedestal-top electron temperature $T_{\mathrm{e,ped}}$ exceeding a certain value,
and the $T_{\mathrm{e,ped}}$ saturates quickly after the appearance of these modes.
Here, $T_{\mathrm{e,ped}}$ is measured by ECE \cite{Zhaohl} at the normalized radius $\rho_t\simeq0.91$.
Once the RMP shortens the ELM period in P3, magnetic perturbations are reduced to a weak $n=2$ harmonic with greatly reduced amplitude and frequency ($\sim30$ kHz).
All coherent modes disappeared after ELM suppression. 
This comparison helps to identify that the excitation of these modes requires a threshold $T_{\mathrm{e,ped}}$. 
The mode frequency here keeps almost constant suggesting the saturation of pedestal electron pressure gradient (shown later), which is different from the washboard mode previously observed on JET \cite{Perez_2004}. 
It indicates that these low-$n$ modes cause the saturation of pedestal electron temperature and pressure, i.e., they regulate the pedestal pressure profile.

\begin{figure}[ht]
	\centering
	\includegraphics[width=\columnwidth]{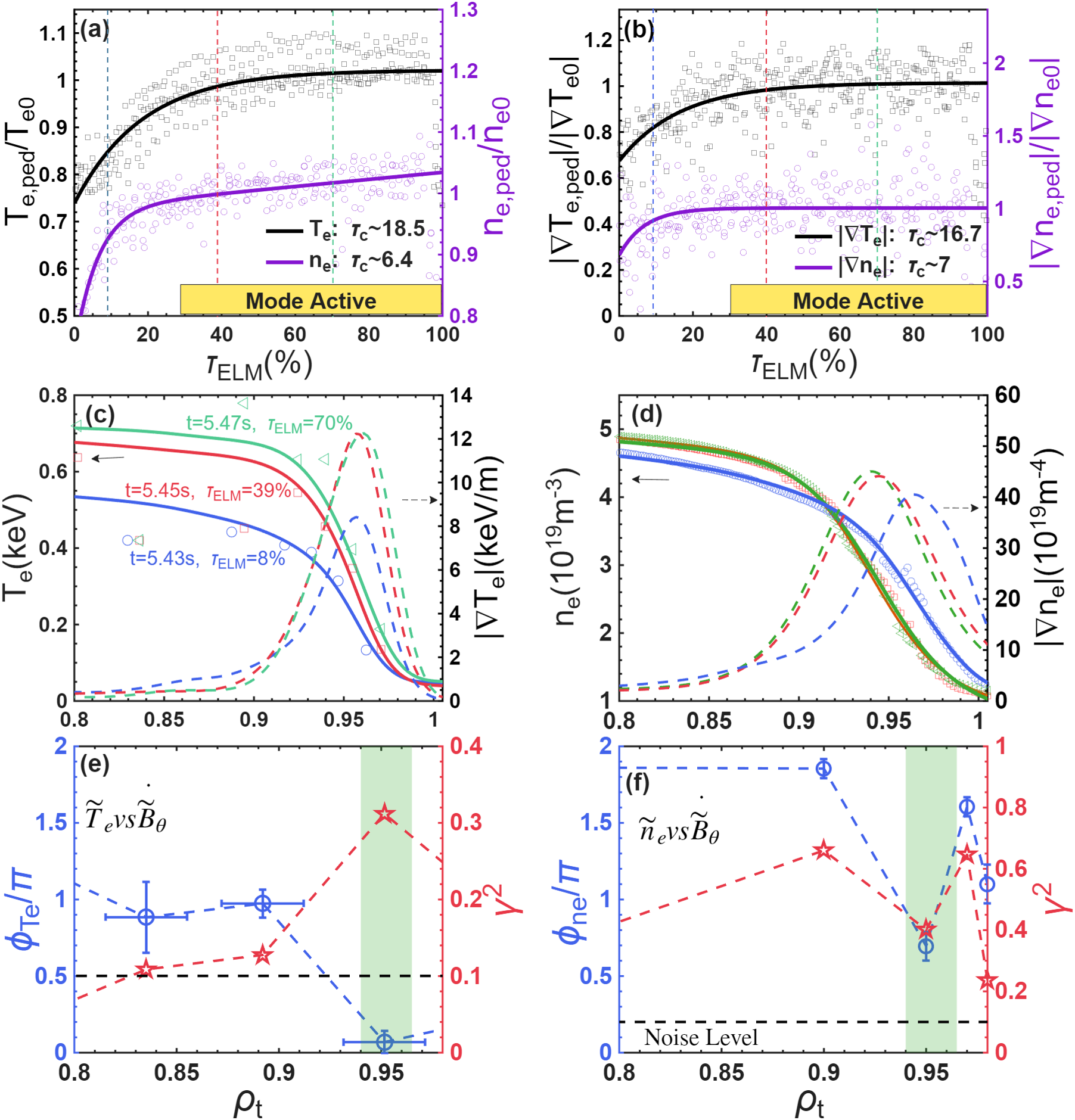}  
	\caption{\label{fig:2}  Statistics of pedestal evolution for multiple ELM cycles on (a) normalized  $T_{\mathrm{e, ped}}/T_{\mathrm{e0}}$ (black squares) and $n_{\mathrm{e, ped}}/n_{\mathrm{e0}}$ (purple circles),  and (b) their radial gradients by aligning their onset time for pedestal recovery, 
		together with radial profiles on pedestal (c)  electron temperature and (d) density (solid) and their gradients (dashed) for three different stages within one ELM cycle at $\tau_{\mathrm{ELM}}=8\%$ (blue), $39\%$ (red) and $70\%$ (green) in the EAST discharge \#155884. Radial profiles of the cross phase $\phi$ (blue circles) and coherence $\gamma^2$ (red stars) between (e) $\tilde{T}_e$ and $\dot{\tilde{B}}_\theta$, (f) $\tilde{n}_e$ and $\dot{\tilde{B}}_\theta$  at $t\simeq6.36$s. The horizontal dashed line marks the coherence noise level, and the green band indicates the mode localization region.   
}
\end{figure}

Statistics of pedestal evolution identify the electron temperature gradient as the mode drive and establish the ``active constraint'', i.e. the two-way constraint between the critical gradient and the mode. 
Figs.~\ref{fig:2}(a) and (b) show the statistical results of pedestal recovery in multiple ELM cycles in another discharge with similar inter-ELM magnetic fluctuation behavior. 
 Here, the normalized ELM-cycle time is defined as $\tau_{\mathrm{ELM}}(\%)=100\%\times\frac{t-t_{\mathrm{crash}}(\mathrm{i-1})}{t_{\mathrm{crash}}(\mathrm{i})-t_{\mathrm{crash}}(\mathrm{i-1})}$ for the $i^{\mathrm{th}}$ ELM cycle.
 Solid curves are fitted to $y(\tau) = y_0 + A\left(1 - e^{-\tau/\tau_c}\right)$.
The density pedestal height and gradient saturate within $\tau_{\mathrm{ELM}} \sim10\%$ of the cycle, which is much earlier than the modes' onset.
They remain almost unchanged, indicating that the particle channel is essentially unaffected after the appearance of these modes. 
The electron temperature channel, by contrast, exhibits a fully self-consistent sequence. 
Once the temperature gradient reaches the critical value, the mode is excited almost instantaneously, on a timescale far shorter than the transport timescale.
This means that the modes are driven by the electron temperature gradient. 
The saturation of pedestal electron temperature and hence the pressure gradient after the appearance of the modes is consistent with the constant mode frequencies discussed earlier. 
The convergence of different ELM cycles to the same saturated gradient is the hallmark of a critical-gradient constraint.
This means that the pedestal electron pressure gradient is indeed clamped by these modes.



The pedestal electron temperature and density profiles in Figs.~\ref{fig:2}(c) and (d), taken at $\tau_{\mathrm{ELM}}=8\%$, $39\%$ and $70\%$, confirm this sequence. 
Here, the electron temperature is measured by Thomson scattering (TS) diagnostics \cite{ZangQ_2016} and the density is measured by reflectometry \cite{QuH_2015}. 
The density profiles are averaged over a $\pm2.5$ ms time window.
Between the first two time slices, prior to the mode onset, the $T_e$ gradient steepens markedly, 
while the $n_e$ gradient changes little apart from a radial shift of its peak, which further identifies the pedestal electon temperature as the free-energy source of the instability. 
Between the second and third instants, across the mode-active phase, both gradient profiles remain nearly unchanged while only the pedestal top $T_e$ still rises slowly, which is the profile stiffness expected of a clamped critical gradient. 
This morphology differs from previous pedestal reports, where the pressure profile widens along a nearly fixed boundary \cite{Dickinson_2012_PRL,Groebner_2010,Burckhart_2010}.
Here, the clamped $T_e$ profile remains essentially fixed while the density pedestal keeps expanding inward at an early stage.

\begin{figure}[t]
	\centering
	\includegraphics[width=\columnwidth]{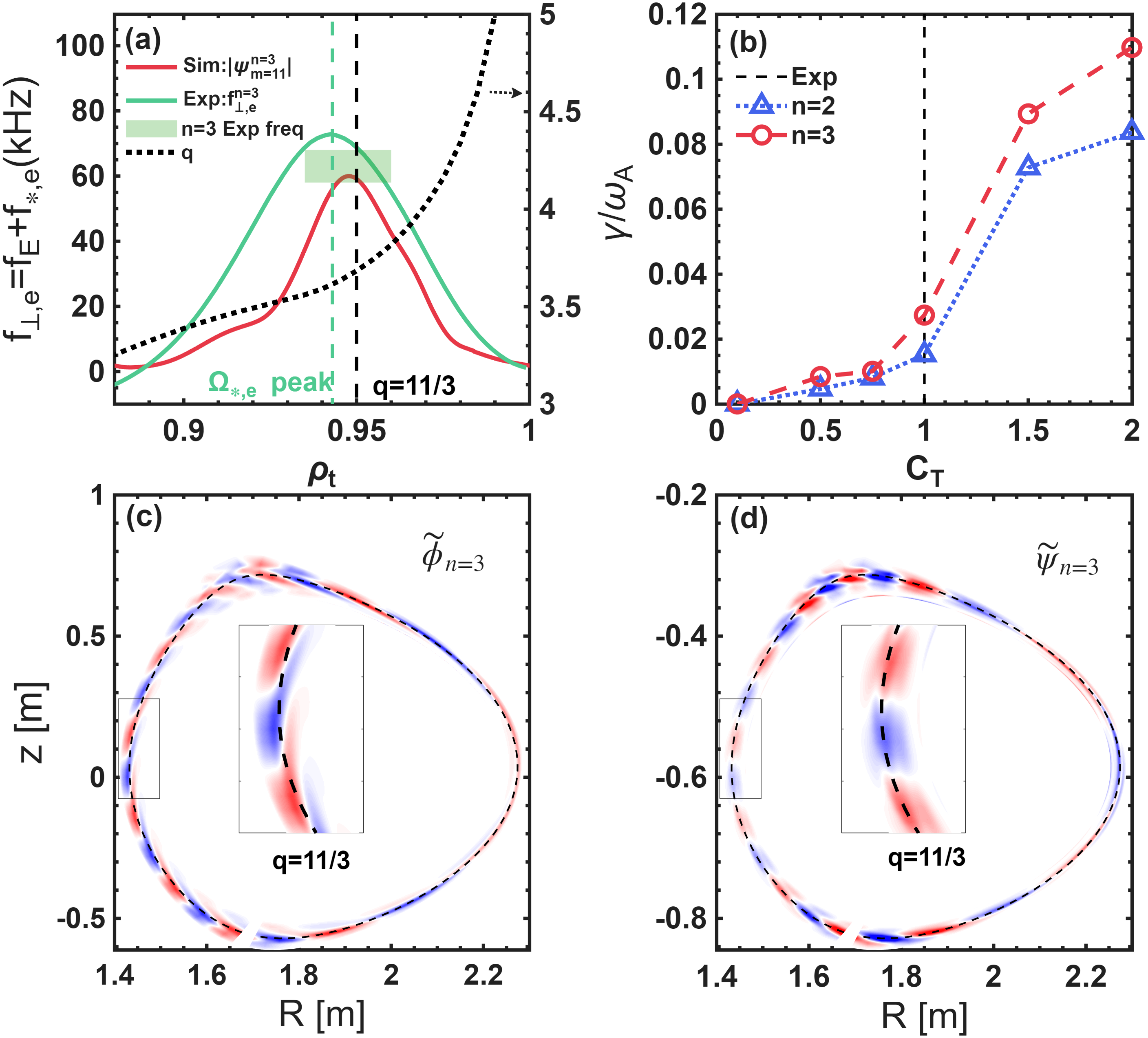}
	\caption{\label{fig:4} 
		Experimental and simulated identification of the inter-ELM fluctuations as MTMs. (a) Pedestal-region profiles for discharge \#80655: the $n=3$ electron perpendicular rotation frequency $3 f_{\perp,e}$ (green solid), the safety factor $q$ (black dotted), and the distribution of the $m/n=11/3$ Fourier component $|\tilde{\psi}_{m=11}^{n=3}|$ (red solid). The vertical black (green) dashed line marks the $q=11/3$ resonant surface (the peak of $\Omega_{\ast,e}$), and the green band denotes the measured $n=3$ magnetic fluctuation frequency. (b) Linear growth rate $\gamma/\omega_A$ versus the temperature gradient coefficient $C_T$ for $n=2$ (blue triangles) and $n=3$ (red circles); the vertical dashed line marks the experimental value of $C_T$. (c) and (d) Poloidal mode structures of the $n=3$ $\tilde{\phi}$ and $\tilde{\psi}$ in the $R$--$Z$ plane, where the black dashed lines trace the $q=11/3$ resonant surface.
	}
\end{figure}

The measured tearing parity and the match of the mode frequency to the electron diamagnetic drift provide experimental confidence that the modes are MTMs. 
As shown in Figs.~\ref{fig:2}(e) and (f), the coherence of both temperature perturbation $\tilde{T}_e$ and density perturbation $\tilde{n}_e$ with the magnetic perturbation $\dot{\tilde{B}}_\theta$ peaks inside the pedestal gradient region (green band), well above the noise level, demonstrating that the modes are localized at the location where the electron temperature gradient is steepest.
Across the same region, both cross phases exhibit a pronounced jump of $\Delta\phi\sim\pi$, indicating a signature of tearing parity. 
Taking the $n=3$ harmonic as an example, the observed $n=3$ mode frequency by the magnetic probes (green band) is very close to the perpendicular rotation frequency $3 f_{\perp,e}= 3\left( f_E+f_{\ast,e}\right) $ at the $q=11/3$ resonant surface, as shown in Fig.~\ref{fig:4}(a).
The frequency match occurs near both the $q=11/3$ resonant surface and the maximum of the $\Omega_{\ast,e}$ profile. 
Similar matching is observed for other harmonics. 
This is consistent with the location alignment requirement for MTM excitation found in previous simulations \cite{larakers2021,Curie2022pop,Hatch_2021}.

Global fluid simulations confirm the MTM identification and reveal the threshold behavior that makes the critical-gradient constraint self-limiting.
We perform global fluid simulations with the four-field model of \cite{Fank2024} in the BOUT++ framework \cite{Dudson2009,Xuxq2010}, which captures the key MTM physics. 
The electron parallel momentum equation,
\begin{eqnarray}
	n_em_e\alpha^{\prime\prime}\frac{\partial}{\partial t}V_{e\parallel}&=&en_e\left(\nabla_{\parallel}\phi+\frac{\partial}{\partial t}A_{\parallel}\right)-\nabla_{\parallel}p_{e}-\alpha n_e\nabla_{\parallel}T_{e} \nonumber\\
	&& -(en_e)^{2}\eta V_{e\parallel}+\frac{C_T\alpha\alpha^{\prime}n_e}{C_\nu\nu_{ei}}\frac{\partial}{\partial t}\left(\nabla_{\parallel}T_{e}\right)
\end{eqnarray}
retains the time-dependent thermal force (last term), which provides the $\tilde{T}_e$--$\tilde{\psi}$ phase shift required for the $\nabla T_e$ drive of tearing-parity modes, i.e., the essential collisional MTM physics \cite{Hassam1980_tear,Hassam1980}. Other coefficients are set to $\alpha=0.8, \alpha^\prime=0.54, \alpha^{\prime\prime}=1.24$ \cite{Hassam1980_tear}. $C_T$ and $C_\nu$ are scaling coefficients, $\nu_{ei}$ is the electron-ion collision frequency.
Scanning the thermal-force coefficient $C_T$ (nominally unity) in Fig.~\ref{fig:4}(b), both $n=2$ and $n=3$ are stabilized as $C_T\rightarrow0$ and destabilized with increasing drive, identifying $\partial_t(\nabla_{\parallel}T_e)$ as the drive. The scan exhibits a clear threshold behavior near the experimental gradient: the growth rates remain small up to $C_T\simeq1$ and rise sharply beyond, indicating that the instability is highly sensitive to the thermal-force drive around the experimental value. This threshold makes the constraint self-limiting: any further increase of pedestal $T_e$ gradient beyond the critical value sharply strengthens the MTM drive, and the resulting electron heat flux, at experimentally relevant levels \cite{Hatch_2021,Doerk2011}, flattens the local gradient back toward the critical value.

The simulated mode structures support this identification: the dominant poloidal Fourier component $|\tilde{\psi}_{m=11}^{n=3}|$ (red curve in Fig.~\ref{fig:4}(a)) is radially localized between the $\Omega_{\ast,e}$ maximum and the $q=11/3$ resonant surface, consistent with slab MTM alignment theory \cite{larakers2021,Hatch_2021}. The same holds for the $n=2$ component (not shown). $\tilde{\phi}$ and $\tilde{\psi}$ display the odd and even tearing parity mode structure (Figs.~\ref{fig:4}(c) and (d)), and $\tilde{\psi}$ exhibits a non-ballooning structure concentrated towards the top and bottom of the tokamak \cite{Hassan_2022,Hatch_2016}. Together with the electron-diamagnetic propagation, an ion-scale wavelength and $\nabla T_e$-driven clamping, these results identify the inter-ELM fluctuations as MTMs.

\begin{figure}[t]
	\centering
	\includegraphics[width=0.65\columnwidth]{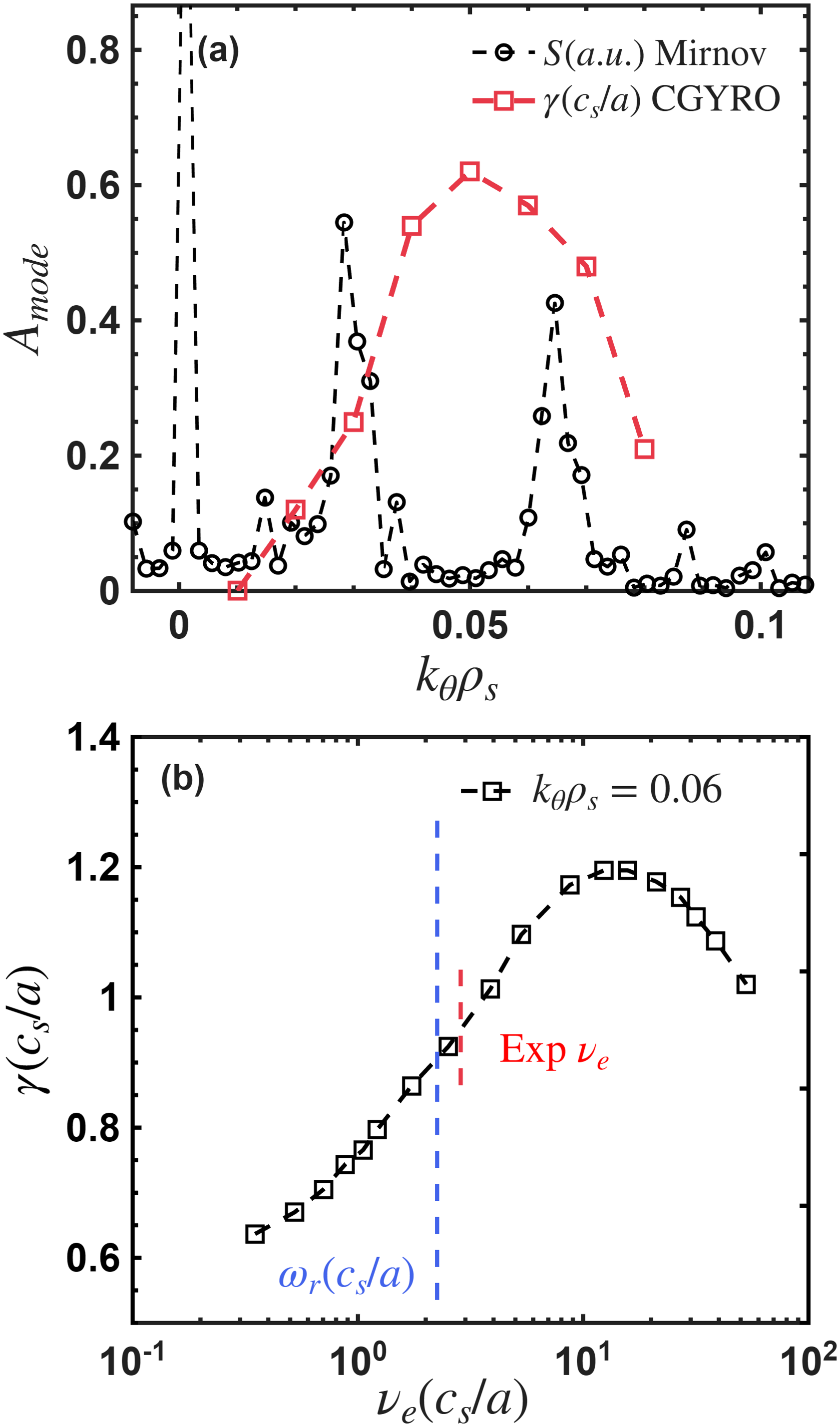}
	\caption{\label{fig:5} (a) Poloidal wavenumber spectrum $S(k_\theta\rho_s)=\sum_{f}^{}S(k_\theta,f)$ of the mode (black circles), estimated from a pair of poloidally separated low-field-side Mirnov probes near the midplane during $t=3959$--$3964$ ms of discharge \#80652, compared with the CGYRO linear growth rate $\gamma$ (red squares) as a function of $k_\theta\rho_s$ at $\rho=0.95$. (b) CGYRO growth rate at $k_\theta\rho_s=0.06$ versus collisiona frequency $\nu_e/(c_s/a)$; the red (blue) vertical dashed line marks the experimental collisionality (the collisionality at which the simulated real frequency matches the measured mode frequency).}
\end{figure}

Gyrokinetic simulations show that the MTM is unstable in the regime $\omega_{MTM}\sim\omega_{\ast,e}\sim\nu_{ei}$, which is commonly satisfied in modern tokamak pedestals.
Using the gyrokinetic code CGYRO \cite{Candy2016CGYRO,Jian2019PRL}, the simulated growth rate spectrum peaks at $k_\theta\rho_s\simeq0.05$, between the two measured peaks of the poloidal wavenumber spectrum $S(k_\theta\rho_s)$ at $\simeq0.03$ and $\simeq0.06$, with the unstable band overlapping both, as shown in Fig.~\ref{fig:5}(a). 
This reproduces the measured ion-scale spectral range. 
Its collisionality dependence is non-monotonic [Fig.~\ref{fig:5}(b)], a hallmark of MTMs \cite{Drake1977,larakers2021}, with the experimental collisionality lying close to the growth rate maximum.
In the fluid simulation, where the fluid closure is derived in the strongly collisional regime, collisionality is instead found to be stabilizing, a trend opposite to that of classical tearing modes. 
At the experimental parameters, the mode frequency, electron diamagnetic frequency and collision frequency are all of the same order, $\omega_{MTM}\sim\omega_{\ast,e}\sim\nu_{ei}$, which is the characteristic ordering of collisional slab-branch MTMs.

This recovery picture complements, rather than contradicts, the EPED model \cite{Snyder_2009}.
A gyrokinetic check indicates that KBMs are second-stable in the MTM-active mid-pedestal region due to large $\alpha$ and locally flattened magnetic shear $\hat{s}$, consistent with recent findings \cite{Hatch2026pop}. MTMs therefore provide an additional gradient constraint on the pedestal pressure profile, while KBMs and drift-type instabilities may remain important in the pedestal top and foot. The detailed pedestal stability analysis will be presented in a forthcoming work.

In summary, the regulation of the pedestal pressure profile by MTMs in the EAST tokamak is presented. 
Supported by global fluid and local gyrokinetic simulations, this first evidence that MTMs regulate the pedestal pressure profile through a critical-gradient mechanism in ITER-baseline-like plasmas is potentially important for understanding pedestal formation.
MTMs arrest the recovery of the electron temperature pedestal height by clamping its gradient at a critical value, while leaving the density profile essentially unaffected. The process is self-limiting from both sides. 
Above the critical gradient, the sharply enhanced thermal-force drive produces strong MTM turbulence whose electron heat flux is expected to flatten the local gradient back toward the critical value; below it, the weakened modes hardly affect the pedestal recovery, as exemplified by the RMP phase, where mitigated ELMs prevent the gradient from reaching the strong-excitation level, so that the surviving weak $n=2$ harmonic has negligible influence on the $T_e$ recovery. 
These low-$n$ modes rotate in the electron diamagnetic direction at frequencies close to the electron diamagnetic frequency, exhibit a radial tearing-parity structure, and carry significant electron heat transport, dominating the inter-ELM pedestal evolution. 
They also help explain the different recovery timescales of the density and temperature pedestals after an ELM crash.   
Therefore, electromagnetic instabilities play a defining role in the H-mode pedestal: not only do KBMs limit pedestal evolution, but MTMs also impose an additional constraint on determining pedestal structure in tokamaks, providing a physics basis for predictive pedestal models in ITER-baseline scenarios. 
However, self-consistent modeling of the nonlinear MTM evolution and the resulting pedestal profile saturation remains challenging. 
A more comprehensive assessment of the pedestal instabilities and the associated transport is needed, toward a quantitative understanding of the inter-ELM pedestal dynamics.

\begin{acknowledgments}
This work was supported by the National MCF Energy R\&D Program of China under Grants No. 2024YFE03010000, and the Strategic Priority Research Program of Chinese Academy of Sciences under Grant No. XDB0790100. 
\end{acknowledgments}

\bibliography{ref.bib}

\end{document}